\documentclass[twocolumn,showpacs,amsmath,amssymb,prl]{revtex4}
\usepackage{graphicx}
\usepackage{dcolumn}
\usepackage{bm}
\usepackage{psfig}
\newcommand{\kskl}{K_S^0 K_L^0}
\newcommand{\ks}{K_S^0}
\newcommand{\kl}{K_L^0}
\newcommand{\BR}{{\cal B}}
\newcommand{\eff}{\varepsilon}

\newcommand{\psp}{\psi(2S)}
\newcommand{\jpsi}{J/\psi}
\newcommand{\chicJ}{\chi_{cJ}}

\newcommand{\chico}{\chi_{c1}}

\newcommand{\EE}{e^+e^-}
\newcommand{\MM}{\mu^+\mu^-}

\newcommand{\piz}{\pi^0}
\newcommand{\pp}{\pi^+\pi^-}
\newcommand{\kk}{K^+K^-}
\newcommand{\ppb}{p\overline{p}}

\newcommand{\jpsipp}{\pi^+\pi^-J/\psi}

\newcommand{\rhopi}{\rho\pi}
\newcommand{\ra}{\rightarrow}
\newcommand{\jpsito}{J/\psi \rightarrow }

\newcommand{\pspto}{\psi(2S) \rightarrow }
\newcommand{\chicJto}{\chi_{cJ} \rightarrow }

\newcommand{\chicoto}{\chi_{c1} \rightarrow }

\newcommand{\bfg}{\begin{figure}}
\newcommand{\efg}{\end{figure}}
\newcommand{\bitm}{\begin{itemize}}
\newcommand{\eitm}{\end{itemize}}
\newcommand{\bnum}{\begin{enumerate}}
\newcommand{\enum}{\end{enumerate}}
\newcommand{\btbl}{\begin{table}}
\newcommand{\etbl}{\end{table}}
\newcommand{\btbu}{\begin{tabular}}
\newcommand{\etbu}{\end{tabular}}
\begin{document}

\preprint{Draft-PRL}

\title{\boldmath First observation of $\pspto \kskl$}
\author{
J.~Z.~Bai$^1$,        Y.~Ban$^{10}$,         J.~G.~Bian$^1$,
X.~Cai$^{1}$,         J.~F.~Chang$^1$,       H.~F.~Chen$^{16}$,    
H.~S.~Chen$^1$,       H.~X.~Chen$^{1}$,      J.~Chen$^{1}$,        
J.~C.~Chen$^1$
% ...
\\(BES Collaboration)\\ 
$^1$ Institute of High Energy Physics, Beijing 100039, People's Republic of
     China\\
$^2$ China Center of Advanced Science and Technology, Beijing 100080,
     People's Republic of China\\
$^3$ Guangxi Normal University, Guilin 541004, People's Republic of China\\
$^4$ Henan Normal University, Xinxiang 453002, People's Republic of China\\
$^5$ Huazhong Normal University, Wuhan 430079, People's Republic of China\\
$^6$ Hunan University, Changsha 410082, People's Republic of China\\                                                  
$^7$ Liaoning University, Shenyang 110036, People's Republic of China\\
$^{8}$ Nanjing Normal University, Nanjing 210097, People's Republic of China\\
$^{9}$ Nankai University, Tianjin 300071, People's Republic of China\\
$^{10}$ Peking University, Beijing 100871, People's Republic of China\\
$^{11}$ Shandong University, Jinan 250100, People's Republic of China\\
$^{12}$ Shanghai Jiaotong University, Shanghai 200030, 
        People's Republic of China\\
$^{13}$ Sichuan University, Chengdu 610064,
        People's Republic of China\\                                    
$^{14}$ Tsinghua University, Beijing 100084, 
        People's Republic of China\\
$^{15}$ University of Hawaii, Honolulu, Hawaii 96822\\
$^{16}$ University of Science and Technology of China, Hefei 230026,
        People's Republic of China\\
$^{17}$ Wuhan University, Wuhan 430072, People's Republic of China\\
$^{18}$ Zhejiang University, Hangzhou 310028, People's Republic of China
}

\date{\today}
           
\begin{abstract}

The decay $\pspto \kskl$ is observed for the first time using $\psp$
data collected with the Beijing Spectrometer (BESII) at the Beijing
Electron Positron Collider (BEPC); the branching ratio is determined
to be \( \BR(\pspto \kskl) = (5.24\pm 0.47 \pm 0.48)\times 10^{-5}\).
Compared with $\jpsito \kskl$, the $\psp$ branching ratio is enhanced
relative to the prediction of the perturbative QCD ``12\%'' rule.  The
result, together with the branching ratios of $\psp$ decays to other
pseudoscalar meson pairs ($\pp$ and $\kk$), is used to investigate the
relative phase between the three-gluon and the one-photon annihilation
amplitudes of $\psp$ decays.

\end{abstract}


\maketitle


It has been determined that for many two-body exclusive $\jpsi$
decays, like vector pseudoscalar (VP), vector vector (VV), and
pseudoscalar pseudoscalar (PP) meson decays and nucleon anti-nucleon
(N$\overline{\hbox{N}}$) decays, the relative phases between the
three-gluon and the one-photon annihilation amplitudes are near
$90^\circ$~\cite{suzuki,dm2exp,mk3exp,a00,a11,ann}.  For $\psp$
decays, the available information about the phase is much more limited
because there are fewer experimental measurements.  It has been argued
that the relative phases in $\psp$ decays should be similar to those
in $\jpsi$ decays~\cite{suzuki,gerard}, but the analysis of $\psp$ to
VP decays in Ref.~\cite{suzuki} indicates this phase is likely to be
around $180^\circ$. Another analysis of this mode though shows the
relative phase observed in $\jpsi$ decays could also fit these
decays~\cite{wymphase}, but it could not rule out the $180^\circ$
possibility due to the big uncertainties in the experimental
data. Therefore it is important to measure phases in other $\psp$
decay modes.

In $\pspto \hbox{PP}$, the currently available measurements on $\pp$ and
$\kk$ are from DASP~\cite{dasp}, with huge errors.  These do not
provide enough information to extract the phase since there are three
free parameters in the parametrization of the $PP$
amplitudes~\cite{a00,a11,haber}. The result for $\pspto \kskl$ is also
needed
to get the phase~\cite{phase_pp}.

Furthermore, there is a longstanding ``$\rhopi$ puzzle'' between
$\jpsi$ and $\psp$ decays in some decay modes; many $\psp$ decay channels
compared with the corresponding $\jpsi$ decays are suppressed relative
to the perturbative QCD predicted ``12\% rule''~\cite{beswangwf}.  It
is of great interest to check this in more channels.


In this letter, the first observation of $\pspto \kskl$ is reported,
and its branching ratio is used to determine the phase between the
three-gluon and one-photon annihilation amplitudes and to test the
``12\% rule'' between $\jpsi$ and $\psp$ decays. The data used for the
analysis are taken with the Beijing Spectrometer (BESII) detector at
the Beijing Electron-Positron Collider (BEPC) storage ring at a
center-of-mass energy corresponding to $M_{\psp}$.  The data sample
contains a total of $14(1 \pm 5\%)\times 10^6$ $\psp$ decays, as
determined from inclusive $\psp$ hadronic decays~\cite{pspscan}.

BES is a conventional solenoidal magnet detector that is
described in detail in Ref.~\cite{bes}, BESII is the upgraded version
of the BES detector~\cite{bes2}. A 12-layer vertex
chamber (VC) surrounding the beam pipe provides trigger
information. A forty-layer main drift chamber (MDC), located
radially outside the VC, provides trajectory and energy loss
($dE/dx$) information for charged tracks over $85\%$ of the
total solid angle.  The momentum resolution is
$\sigma _p/p = 0.017 \sqrt{1+p^2}$ ($p$ in $\hbox{\rm GeV}/c$),
and the $dE/dx$ resolution for hadron tracks is $\sim 8\%$.
An array of 48 scintillation counters surrounding the MDC  measures
the time-of-flight (TOF) of charged tracks with a resolution of
$\sim 200$ ps for hadrons.  Radially outside the TOF system is a 12
radiation length, lead-gas barrel shower counter (BSC).  This
measures the energies
of electrons and photons over $\sim 80\%$ of the total solid
angle with an energy resolution of $\sigma_E/E=22\%/\sqrt{E}$ ($E$
in GeV).  Outside of the solenoidal coil, which
provides a 0.4~Tesla magnetic field over the tracking volume,
is an iron flux return that is instrumented with
three double layers of  counters that
identify muons with momentum greater than 0.5~GeV/$c$.


A Monte Carlo simulation is used for the determination of the
detection efficiency. For the signal channel, $\pspto \kskl$, the
angular distribution of the $\ks$ or $\kl$ is generated as
$\sin^2\theta$, where $\theta$ is the polar angle in the laboratory
system. The $\kl$ is allowed to decay and interact with the material
in the detector, and for the $\ks$, only $\ks \ra \pp$ is generated.
The detector response is simulated using a Geant3 based Monte Carlo
program, SIMBES. Reasonable agreement between data and Monte Carlo
simulation has been observed in various channels tested, including
$\EE \ra (\gamma)\EE$, $\EE\ra (\gamma)\MM$, $\jpsito \ppb$ and
$\pspto \jpsipp, \jpsito \ell^+\ell^-$ $(\ell=e,\mu)$.


Candidate events are 
required to satisfy the following selection criteria:
\begin{enumerate}
\item   The number of charged tracks is required to be two
        with net charge zero.

\item   Each track should satisfy
        $|\cos\theta|<0.8$, where $\theta$ is the polar angle
        in the MDC, and should have good helix
        fit so that the error matrix from track fitting is
        available for secondary vertex finding.

\item   $E_{\gamma}^{tot}<1.0$~GeV, where $E_{\gamma}^{tot}$ is the
        total energy of the neutral clusters in the BSC which are not
        associated with the charged tracks. 
\end{enumerate}

\begin{figure}[htbp]
\centerline{\hbox{
\psfig{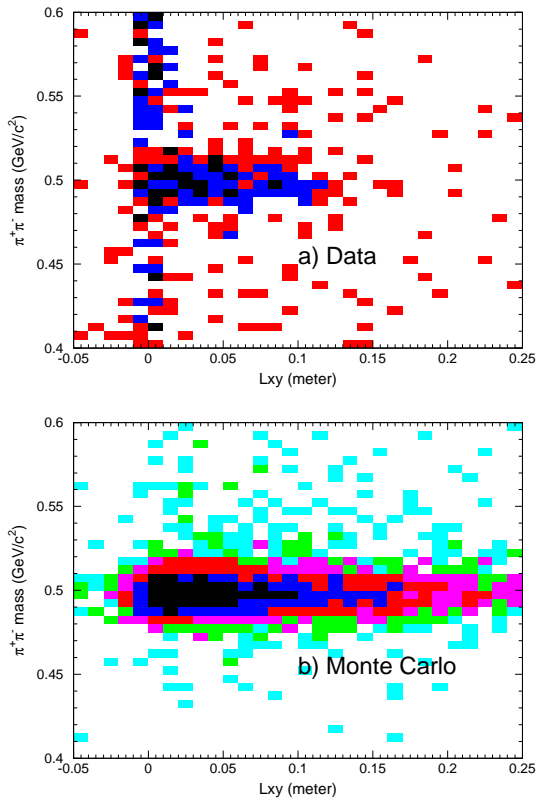}}}
\caption{Scatter plot of $\pp$ invariant mass versus the decay 
length in the transverse plane for events with $\pp$ momentum greater 
than 1.7~GeV/$c$ for a) data and b) Monte Carlo simulation.}
\label{mkslxy}
\end{figure}

The two tracks are assumed to be $\pi^+$ and $\pi^-$.  To find the
intersection of the two tracks, an iterative, nonlinear least squares
technique is used.  The intersection is taken as the $\ks$ vertex, and
the momentum of the $\ks$ is calculated at this point.
Figure~\ref{mkslxy} shows a scatter plot of the $\pp$ invariant mass
versus the decay length in the transverse plane ($L_{xy}$) for events
that satisfy the above selection criteria and have $\pp$ momentum
greater than 1.7~GeV/$c$. The cluster of events with mass consistent
with the nominal $\ks$ mass and with a long decay length indicates a
clear $\ks$ signal. The lack of events at $L_{xy}>0.1$~m is due to the
trigger, which will be discussed later.

\begin{figure}[htbp]
\centerline{\hbox{
\psfig{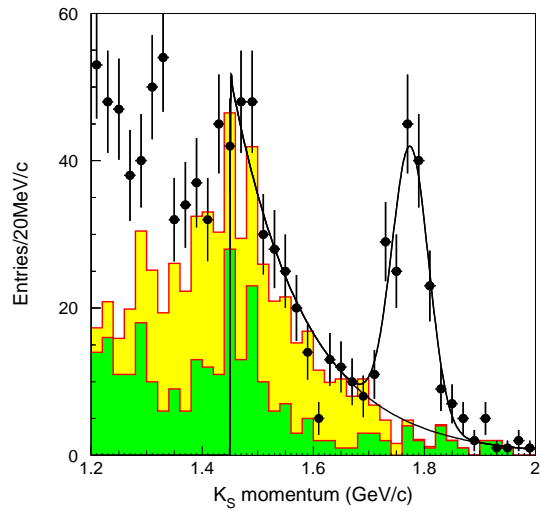}}}
\caption{The $\ks$ momentum distribution. The dots
with error bars are data, the dark shaded histogram is from 
$\ks$ mass side band events, and the light shaded histogram is from the 
Monte Carlo simulated backgrounds. The curve shown 
in the plot is the best fit of the data.}
\label{pksdtfit}
\end{figure}

Fits of the $\pp$ invariant mass distributions (not shown) indicate good 
agreement between data and Monte Carlo simulation in both mass and 
the mass resolution. After requiring $L_{xy}>1$~cm and $\pp$ mass within 
twice the mass resolution around the $\ks$ nominal mass,
the $\ks$ momentum distribution, shown in
Figure~\ref{pksdtfit}, is obtained. In the plot, there is
a clear peak at around 1.77~GeV/$c$ with low 
background, as indicated by the $\ks$ mass 
side band (three sigma away from the $\ks$ nominal
mass on both sides) events. The excess at lower 
momentum, which is not explained by the side band background,
is due to the contribution of the background channels 
with $\ks$ production. 

The main  $\ks$ production backgrounds near the signal 
region are from $\pspto K^{*0}(892)\overline{K^0} + c.c.$ and
$\pspto \gamma \chico, \chicoto K^{*0}(892)\overline{K^0} + c.c.$ 
where the $K^{*0}$ decays into $K^0$ and $\piz$ and one of the
$K^0$s becomes a $\ks$ and the other one becomes a $\kl$. The background
from $\pspto \gamma \chicJ (J=0,2), \chicJto \ks\ks$ where one of the  
$\ks$ decays into $\pp$ and the other decays into $\piz\piz$
is removed almost entirely (more than 95\%) by the $E_{\gamma}^{tot}$
cut, according to the Monte Carlo simulation. The decay
$\pspto \jpsi+X$ with $\jpsi$ decaying into $\ks X$ has a big
branching ratio, but the $\ks$ momentum is much lower.
The light shaded histogram in Figure~\ref{pksdtfit} shows the 
contribution of the background channels normalized to the known
branching ratios~\cite{beschic,pdg}. It can be seen that the agreement 
between the background estimation and data is good near the $\kskl$ peak,
indicating that the estimation of the background under the $\kskl$ peak is
reliable. 

\begin{figure}[htbp]
\centerline{\hbox{
\psfig{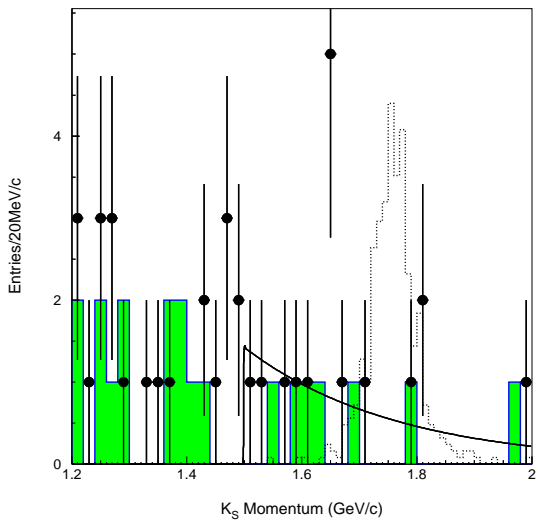}}}
\caption{The $\ks$ momentum distribution for data at
$\sqrt{s}=3.65$~GeV. The dots with error bars are data; the shaded
histogram is for $\ks$ mass side band events.  The curve shown in the
plot is the best fit of the data, while the dotted histogram is the
expected shape of a $\kskl$ signal as determined by Monte Carlo
simulation (not normalized).}
\label{pksfit365}
\end{figure}

Under SU(3) symmetry, $\kskl$ production via virtual photon
annihilation is forbidden.  This is checked by applying the same
selection criteria to the data sample taken below the $\psp$ peak, at
$\sqrt{s}=3.650$~GeV, with an integrated luminosity of
about one-third of that at the $\psp$ peak.  Figure~\ref{pksfit365} shows
the $\ks$ momentum spectrum of the selected events; the events in the
signal region agree well with expectation from the $\ks$ mass side
band events.  Taking all four candidates with momentum between 1.7 and
1.9~GeV/$c$ as signal, the upper limit of the production cross section
at the 90\% C.~L. is measured to be $\sigma<5.9~\hbox{pb}$.
The background from the continuum contribution is thus neglected 
in the folowing analysis since no evidence for $\kskl$ production via
the
virtual photon process is observed.

The $\ks$ momentum spectrum of the selected events is fitted from 1.45
to 2.0~GeV/$c$ with a Gaussian distribution for the signal and an
exponential for the background using the unbinned maximum likelihood
method. The result is shown in Figure~\ref{pksdtfit}. The backgrounds
from the $\ks$ mass side bands and the simulated background channels
are also shown, and they agree with the fitted background reasonably
well near the signal region, considering the uncertainties in the
global normalization of the background channels. The peak $\ks$
momentum is $(1775.0\pm 3.3)$~MeV/$c$, which is in good agreement with the
Monte Carlo expectation. The momentum resolution also agrees with the Monte
Carlo simulation well.  The fit yields $n^{obs} = 156\pm 14$ events,
and the efficiency for detecting $\pspto \kskl$, with $\ks\to \pp$ is
$\eff_{MC} = (41.59\pm 0.48)\%$ from the Monte Carlo simulation, where
the error is due to the limited statistics of the Monte Carlo sample.


Due to the long decay length of the high momentum $\ks$ particles and the
trigger requirement for hits in the VC, the trigger efficiency of
$\kskl$ events is very different than for normal hadronic
events. Since the trigger system is not included in the Monte Carlo
simulation, the trigger efficiency is measured using $\kskl$ events by
comparing the number of events beyond and within the outer radius of
the VC with what would be expected for an exponential decay, which
yields a trigger efficiency of $\eff_{trig} = (76.0\pm 1.8)\%$. The systematic error
in the branching ratio measurement from this source and 
all other sources are listed in Table.~\ref{sys}.


The efficiency of the secondary vertex finding, $\eff_{2nd}$, is studied using 
$\jpsito K^{*}(892)\overline{K} + c.c.$ events. It is found that the 
Monte Carlo simulates data fairly well.  Extrapolating the difference 
between data and Monte Carlo simulation to the $\ks$ momentum
range under study and correcting by the polar angle dependence of the
efficiency, a correction factor of $(98.1\pm 4.0)\%$ is obtained
to the Monte Carlo efficiency.


\begin{table}[htbp]
\caption{Summary of systematic errors.}
\begin{center}
\begin{tabular}{l|c}
\hline\hline
Source                      & Systematic errors (\%) \\\hline
MC statistics               &  1.2              \\
Trigger efficiency          &  2.4              \\
Secondary vertex finding     &  4.1              \\
$E_{\gamma}^{tot}$ cut      &  2                \\  
MDC tracking                &  4                \\
Fit range                   &  2.4              \\
Background shape            &  3.0               \\
$N_{\psp}$                  &  5               \\
$\BR(\ks \ra \pp)$          &  0.4              \\
\hline
Total systematic error      &  9.2               \\\hline\hline
\end{tabular}
\end{center}
\label{sys}
\end{table}

The effect of the cut on the total energy of the photon candidates is
checked with $\jpsito \kskl$ events. The efficiency difference between
data and Monte Carlo simulation is measured to be $0.99\pm 0.01$; data
is slightly lower than Monte Carlo simulation.  No correction to the
final efficiency is made, and 2\% is taken as the systematic
error on the efficiency associated with
this cut.

The simulation of the tracking efficiency agrees with data within
1-2\% for each charged track as measured using channels like $\jpsito
\Lambda \overline{\Lambda}$ and $\pspto \pp \jpsi$, $\jpsito \MM$.
The systematic error for the channel of interest is taken
conservatively as 4\% .

The Monte Carlo simulated mass resolution and momentum resolution of
the $\ks$ agree with those determined from data within the statistical
uncertainties. The cut on the $\ks$ mass distributions at 2 standard
deviations introduces a very small systematic bias and is neglected.

Varying the lower and upper bounds of the fitting range results in
a 2.4\% change in the number of the events; using a second order 
polynomial for the background parametrization causes a 3\% change 
in the number of events. These are taken as
systematic errors. The systematic error in the total number of 
$\psp$ events, $N_{\psp}$, which is
measured using inclusive hadrons in the same way as in
Ref.~\cite{pspscan}, is taken as 5\%.  The systematic error on the
branching ratio $\BR(\ks \ra \pp)$ is obtained from the PDG~\cite{pdg}
directly.

\begin{figure}[htbp]
\centerline{\hbox{
\psfig{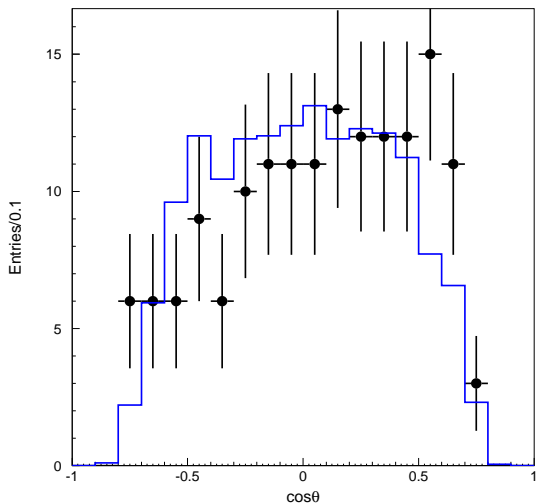}}}
\caption{Distribution of the cosine of the $\ks$ polar angle 
of $\kskl$ events from $\psp$ decays. Dots with error bars are 
data, and the histogram is the Monte Carlo simulation.}
\label{cosks}
\end{figure}

Figure~\ref{cosks} shows the cosine of the $\ks$ polar angle
for $\kskl$ events from $\psp$ decays; agreement
between data and Monte Carlo simulation is observed. 
This distribution is also checked with a larger sample from
$\jpsito \kskl$, where the Monte Carlo simulation agrees with
data very well. This indicates that the angular distribution used 
in the Monte Carlo generator is correct.

The branching ratio of $\pspto \kskl$ is calculated with
\[ \BR(\pspto \kskl)=
    \left.\frac{n^{obs}/(\eff_{MC}\cdot \eff_{trig}\cdot \eff_{2nd})}
               {N_{\psp}\BR(\ks\ra \pp)}
     \right. ~. \]
Using numbers from above (listed in Table.~\ref{br}), one obtains
\[ \BR(\pspto \kskl)
     = (5.24\pm 0.47\pm 0.48)\times 10^{-5} , \]
where the first error is statistical and the second
systematic. 

\begin{table}[htbp]
\caption{Numbers used in the branching ratio calculation and the branching
ratio result.}
\begin{center}
\begin{tabular}{l|c}
\hline\hline
quantity                  & Value \\\hline
$n^{obs}$                 & $156 \pm 14$ \\
$\eff_{MC}$ (\%)          & $41.59 \pm 0.48$ \\
$\eff_{trig}$ (\%)        & $76.0 \pm 1.8$ \\
$\eff_{2nd}$ (\%)         &  $98.1 \pm 4.0$ \\
$N_{\psp} (10^6)$         &  $14.0\pm 0.7$ \\
$\BR(\ks \ra \pp)$        &  $0.6860\pm 0.0027$ \cite{pdg} \\\hline
$\BR(\pspto \kskl) (10^{-5})$ & $5.24\pm 0.47\pm 0.48$  \\
\hline\hline
\end{tabular}
\end{center}
\label{br}
\end{table}

Comparing with the corresponding branching ratio for $\jpsito \kskl$
from BESII ($(1.82\pm 0.14)\times 10^{-4}$)~\cite{bes2ksklj}, one gets
      \[ Q_h = \frac{\BR(\pspto \kskl)}{\BR(\jpsito \kskl)}
                   = (28.8\pm 3.7)\%, \]
where the common errors in the $\jpsi$ and $\psp$ analyses are
removed in the calculation. The result indicates that
$\psp$ decays are enhanced by more than 4$\sigma$ relative to the 
``12\% rule'' expected from pQCD, while for almost all 
other channels where the deviations from the ``12\% rule'' are 
observed, $\psp$ decays are suppressed.

The branching ratio of $\kskl$, together with branching ratios of $\pspto
\pp$ and $\pspto \kk$, can be used to extract the relative phase
between the three-gluon and the one-photon annihilation amplitudes of
the $\psp$ decays to pseudoscalar meson pairs.
It is found that a relative phase around $\pm 90^{\circ}$ can explain 
the experimental results~\cite{phase_pp}.


In summary, the flavor SU(3) breaking process $\kskl$ is observed for
the first time in $\psp$ decays with the BESII $\psp$ data sample, and
the branching ratio is determined to be
\( \BR(\pspto \kskl)
     = (5.24\pm 0.47\pm 0.48)\times 10^{-5} \).
Compared with 
the branching ratio of $\jpsito \kskl$, $\psp$ decays are enhanced
relative to the ``12\%'' pQCD prediction. The phases of the three-gluon and 
the one-photon annihilation amplitudes of $\psp$ decays to 
pseudoscalar meson pairs are found to be nearly orthogonal.


   The BES collaboration thanks the staff of BEPC for their 
hard efforts. This work is supported in part by the National 
Natural Science Foundation of China under contracts 
Nos. 19991480, 10225524, 10225525, the Chinese Academy
of Sciences under contract No. KJ 95T-03, the 100 Talents 
Program of CAS under Contract Nos. U-11, U-24, U-25, and 
the Knowledge Innovation Project of CAS under Contract 
Nos. U-602, U-34(IHEP); by the National Natural Science
Foundation of China under Contract No. 10175060 (USTC); 
and by the Department of Energy under Contract 
No. DE-FG03-94ER40833 (U Hawaii).

\end{document}